\documentclass[10pt,preprint]{emulateapj}
\usepackage{graphics}
\usepackage{subfigure}

\newcommand{\CH}[1]{\colhead{#1}}

\begin{document}
\submitted{Accepted for publication in the Astronomical Journal 25 April 2013
(UMN--TH--3126/12, FTPI--MINN--12/37)} 

\shortauthors{Skillman et al.}
\title{ALFALFA Discovery of the Nearby Gas-Rich Dwarf Galaxy Leo~P.  III. An Extremely Metal
Deficient Galaxy\altaffilmark{\dag}}

\author{
Evan D. Skillman\altaffilmark{1}, 
John J. Salzer\altaffilmark{2,3},
Danielle A. Berg\altaffilmark{1},
Richard W. Pogge\altaffilmark{4},
Nathalie C. Haurberg\altaffilmark{2,3},
John M. Cannon\altaffilmark{5,3},
Erik Aver\altaffilmark{6},
Keith A. Olive\altaffilmark{1,7},
Riccardo Giovanelli\altaffilmark{8},
Martha P. Haynes\altaffilmark{8},
Elizabeth A.K. Adams\altaffilmark{8},
Kristen B.W. McQuinn\altaffilmark{1},
and
Katherine L. Rhode\altaffilmark{2}
}

\altaffiltext{1}{Minnesota Institute for Astrophysics, School of Physics and Astronomy, University of Minnesota, 116 Church St. SE, Minneapolis, MN 55455; 
skillman@astro.umn.edu,
berg@astro.umn.edu, 
olive@physics.umn.edu,
kmcquinn@astro.umn.edu}
\altaffiltext{2}{Astronomy Department, Indiana University, 727 East 3rd Street, Bloomington, IN 47405; 
slaz@astro.indiana.edu,	
nhaurber@astro.indiana.edu,
rhode@astro.indiana.edu}
\altaffiltext{3}{Visiting Astronomer, Kitt Peak National Observatory,
National Optical Astronomy Observatory, which is operated by the
Association of Universities for Research in Astronomy (AURA)
under cooperative agreement with the National Science Foundation.}
\altaffiltext{4}{Department of Astronomy, The Ohio State University, 140 W 18th Ave, Columbus, OH 43210, USA; 
Center for Cosmology and AstroParticle Physics, The Ohio State University, 191 West Woodruff Avenue, Columbus OH 43210; 
pogge@astronomy.ohio-state.edu}
\altaffiltext{5}{Department of Physics and Astronomy, Macalester College, Saint Paul, MN 55105; 
jcannon@macalester.edu}
\altaffiltext{6}{Department of Physics, Gonzaga University, Spokane, WA 99258; 
aver@gonzaga.edu}
\altaffiltext{7}{William I. Fine Theoretical Physics Institute, School of Physics and Astronomy, University of Minnesota, 116 Church St. SE, Minneapolis, MN 55455}
\altaffiltext{8}{Center for Radiophysics and Space Research, Space Sciences Building, Cornell University, Ithaca, NY 14853; 
riccardo@astro.cornell.edu,
haynes@astro.cornell.edu, 
betsey@astro.cornell.edu}

\altaffiltext{\dag}{Some of the observations reported here were obtained at the LBT Observatory.
The LBT is an international collaboration among institutions in the United States, Italy, and Germany. 
LBT Corporation partners are: The University of Arizona on behalf of the Arizona university system; 
Istituto Nazionale di Astrofisica, Italy; LBT Beteiligungsgesellschaft, Germany, representing the Max-Planck Society, 
the Astrophysical Institute Potsdam, and Heidelberg University; 
The Ohio State University, and The Research Corporation, on behalf of 
The University of Minnesota, 
The University of Notre Dame, 
and The University of Virginia.
}


\begin{abstract}
We present KPNO 4-m and LBT/MODS spectroscopic observations of an \ion{H}{2} region 
in the nearby dwarf irregular galaxy Leo~P discovered recently in the Arecibo ALFALFA survey.
In both observations, we are able to accurately measure the temperature 
sensitive [\ion{O}{3}] $\lambda$4363 line 
and determine a ``direct'' oxygen abundance of 12 + log(O/H) = 7.17 $\pm$ 0.04. 
Thus, Leo~P is an extremely metal deficient (XMD) galaxy, and, indeed, one of the most 
metal deficient star-forming galaxies ever observed.  
For its estimated luminosity, Leo~P is consistent with the 
relationship between luminosity and oxygen abundance seen in nearby dwarf galaxies.
Leo~P shows normal $\alpha$ element abundance ratios (Ne/O, S/O, and Ar/O) when compared to
other XMD galaxies, but elevated N/O, consistent with the ``delayed release'' hypothesis
for N/O abundances. 
We derive a helium mass fraction of 0.2509$^{+0.0184}_{-0.0123}$ 
which compares well with the WMAP + BBN prediction of 0.2483 $\pm$ 0.0002 for
the primordial helium abundance. We suggest that surveys of very low mass
galaxies compete well with emission line galaxy surveys for finding XMD
galaxies.  
It is possible that XMD galaxies may be divided into two classes: the relatively rare XMD emission line galaxies
which are associated with starbursts triggered by infall of low-metallicity gas 
and the more common, relatively quiescent XMD galaxies
like Leo~P, with very low chemical abundances due to their intrinsically small masses.
\end{abstract}



\section{INTRODUCTION}\label{sec:intro}

Metallicity determinations of the interstellar media in galaxies allow us to assess 
the chemical evolutionary status of a galaxy.
In this regard, studies of the lowest metallicity, least chemically evolved
galaxies are of special interest \citep{mateo98, ko00, tolstoy09, mcconnachie12}.
Specifically, they are important for studies of: (1) star formation at low metallicity,
which is critical to understanding star formation in the early universe;
(2) stellar properties of low metallicity massive and intermediate mass stars, 
which provide critical constraints on theories of stellar evolution;
(3) constraints on the early enrichment of the pre-galactic medium; and
(4) studies of the primordial elemental abundances. 

\citet{giovanelli12} have reported the discovery of a nearby dwarf irregular galaxy (AGC 208583 = Leo~P)
as part of the Arecibo Legacy Fast ALFA Survey \citep[ALFALFA,][]{giovanelli05, haynes11}.
The ALFALFA survey is a blind survey in the HI 21cm line covering 7000 square degrees
of high Galactic latitude sky. Leo~P was discovered as part of a program to identify
mini-halo candidates in the ALFALFA survey \citep{giovanelli10, adams13}.
Follow-up broadband optical imaging by \citet{rhode12} clearly demonstrated the presence 
of a resolved stellar population at the location of the ALFALFA HI detection, and  
H$\alpha$ imaging in the same study revealed a strong H$\alpha$ source associated with a bright
central star in Leo~P.

Here we report on optical spectroscopy of the emission line source in Leo~P.
We present results of optical spectroscopy from the Kitt Peak National Observatory (KPNO) 4-m 
and from the Large Binocular Telescope (LBT) in Section~\ref{sec:obs}, and derive \ion{H}{2} 
region physical conditions from the spectra in Section~\ref{sec:lines}.
We derive chemical abundances from each of the two spectra in Section~\ref{sec:abund}.
We then compare the results of these abundance analyses to similar analyses
of low metallicity \ion{H}{2} regions in Section~\ref{sec:results}, and discuss the implications 
for our understanding of metal poor galaxies in Section~\ref{sec:diss}.
Finally, we summarize our conclusions in Section~\ref{sec:conclusion}.


\begin{figure}[h]
\epsscale{1.0}
\plotone{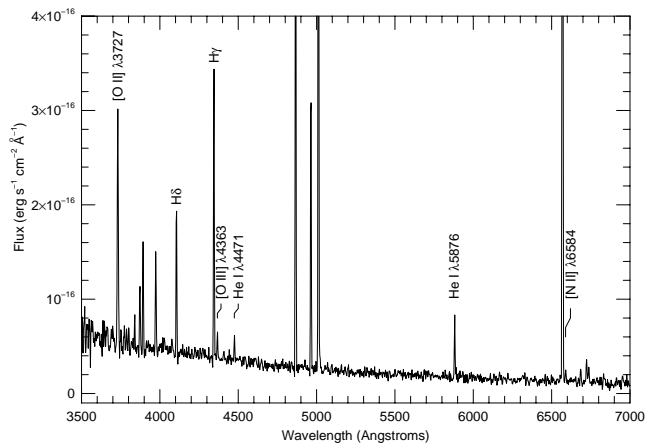}
\caption{The spectrum of the \ion{H}{2} region in Leo~P taken with the KPNO 4-m.
Note the clear detection of the temperature sensitive [\ion{O}{3}] $\lambda$4363
emission line and the very weak strength of the [\ion{N}{2}] $\lambda\lambda$6548,6584 lines,
indicative of a very low abundance.  H$\alpha$, H$\beta$, and \ion{O}{3} $\lambda$5007 are all
off scale to allow better visibility of the weaker emission lines.
}
\label{fig:fig1}
\end{figure}


\section{OPTICAL SPECTROSCOPY OF LEO~P}\label{sec:obs}

\subsection{The KPNO 4-m Spectrum}\label{sec:kpno}

The \ion{H}{2} region discovered by \citet{rhode12} has coordinates of R.A.\ $=$ 10:21:45.1 
and dec.\ $=$ $+$18:05:17.2 (J2000).  A spectrum of the Leo~P \ion{H}{2} region was obtained 
with the KPNO Mayall 4-m on 2012 April 20 under relatively
clear skies and approximately arcsecond seeing.  The Ritchey-Chretien Focus Spectrograph
was used with a Tektronix 2048$^2$ pixel detector (T2KA).  A 316 line mm$^{-1}$ grating
(KPC-10A), $1.5\arcsec$ slit, and WG345 blocking filter were used.  This spectroscopic
configuration yielded a dispersion of 2.78 \AA\ per pixel, a full width at half maximum
resolution of $\lesssim6$ \AA, and a wavelength coverage of 3500--8300 \AA.  Bias frames,
flat-field lamp images, and sky flats were obtained.  Combined helium, argon, and neon arc
lamps were acquired for accurate wavelength calibration.  The spectrophotometric standard
star Feige 34 \citep{mass88} was observed before and after the observations of Leo~P
to yield a secure flux calibration.

Four 1200 second exposures of Leo P were obtained with the slit at a fixed position angle
which approximated the parallactic angle at the midpoint of the observations.  This, in
addition to observing at airmasses of less than $\sim$1.2, served to minimize the
wavelength-dependent light loss due to differential refraction \citep{filippenko82}.

Standard procedures within the IRAF\footnote{IRAF is distributed by the National Optical
Astronomy Observatories, which are operated by the Association of Universities for Research
in Astronomy, Inc., under cooperative agreement with the National Science Foundation.}
\texttt{ccdred} and \texttt{specred} packages were used to bias-subtract, flat-field, and
illumination-correct the raw data frames.   The \texttt{lacos\_spec} routine \citep{vanDokkum01}
was used to remove cosmic rays from the images; this process was checked carefully to ensure
that sharp emission lines in our nebular spectra were not accidentally misidentified as cosmic rays by
the software.  The spectra were then extracted from the 2D images using the IRAF \texttt{apextract}
package.  The individual 1D spectra were wavelength calibrated by applying the solution from
the HeArNe comparison lamps,  and flux-calibrated using the sensitivity curve derived from
the standard star observations.  The latter step also involved making airmass-dependent
atmospheric extinction and reddening corrections using the standard Kitt Peak extinction
curve \citep{crawford70}.  The multiple fully-processed spectra were then combined into
a single composite spectrum.

Figure~\ref{fig:fig1} shows the final calibrated 4-m spectrum.  Note
the clear detection of the temperature sensitive [\ion{O}{3}] $\lambda$4363 emission line which
allows for a ``direct'' abundance determination and the very weak strength of the [\ion{N}{2}]
$\lambda\lambda$6548,6584 lines, indicative of a very low metallicity \citep{dtt02}.


\begin{figure}[h]
\epsscale{1.0}
\plotone{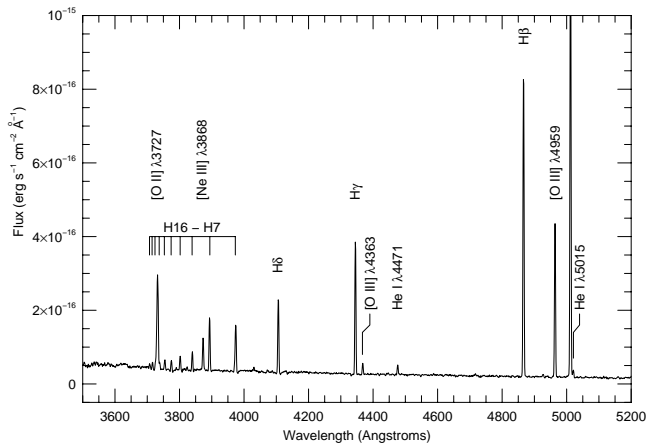}
\caption{The blue portion of the LBT/MODS spectrum of the \ion{H}{2} region in Leo~P.
Note the clear detection of the temperature sensitive [\ion{O}{3}] $\lambda$4363
emission line and the detection of the \ion{He}{1} $\lambda$5015 line.
The higher numbered Balmer series lines are labeled H16 - H7.
Note that underlying absorption is not detected for the Balmer lines.
\ion{O}{3} $\lambda$5007 is off scale to allow better visibility of the weaker emission lines.
}
\label{fig:fig2}
\end{figure}


\subsection{The LBT/MODS Spectrum}\label{sec:lbt}

After the KPNO 4-m spectrum revealed the very low oxygen abundance of Leo~P, we re-observed the
\ion{H}{2} region in Leo~P in order to obtain a higher signal-to-noise spectrum and to detect 
weak emission lines over a larger wavelength range.
A spectrum of the \ion{H}{2} region in Leo~P was obtained with the Multi-Object Double Spectrograph on 
the Large Binocular Telescope \citep[LBT/MODS,][]{pogge10} on 2012 April 29 under relatively clear
skies and approximately 0.6$\arcsec$ seeing.
The LBT/MODS is a double spectrograph with a dichroic that splits the light 
at $\sim$5650 \AA\ and sends it to two separate spectrographs. 
LBT/MODS1 was used in longslit mode with a $1.0\arcsec$ slit imaged onto two 3072 $\times$ 8192 
format e2v CCDs with 15$\mu$m (0.12$\arcsec$) pixels.
The blue side of MODS1 covers a wavelength range of 3200 to 5650 \AA\ with a 400 l mm$^{-1}$ grating providing  a 
resolution of 2.4 \AA .
The red side of MODS1 covers a wavelength range of 5650 to 10000 \AA\ with a 250 l mm$^{-1}$ grating providing  a 
resolution of 3.4 \AA .
Bias frames, flat-field lamp images, and sky flats were obtained.
We observed BD+33$\arcdeg$2642 and Feige 34 as the standard stars with a
5x60\arcsec\  spectrophotometric slit mask near the parallactic
angle.   These are from \citet{oke90}, using the HST CALSPEC flux tables
\citep[e.g.,][]{bohlin10}
which extend further into the UV and Near-IR than the original Oke
fluxes.  Feige 34 was observed at an airmass that more closely matched the airmass of
the target observations, so it was used as a single calibration source instead of being used
in combination with BD+33$\arcdeg$2642.


\begin{figure}[h!]
\epsscale{1.0}
\plotone{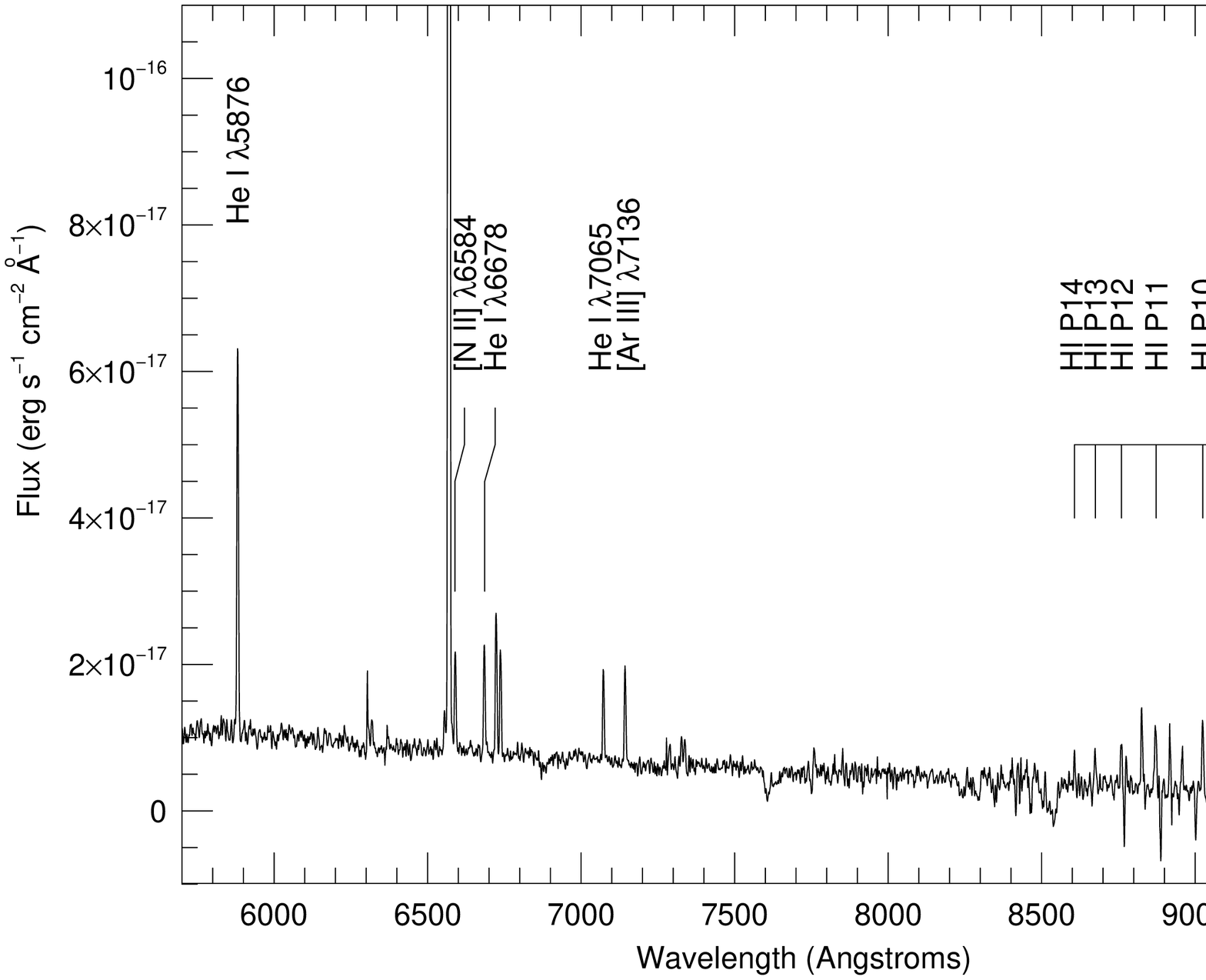}
\caption{The red portion of the LBT/MODS spectrum of the \ion{H}{2} region in Leo~P.
Note the very weak strength of the [\ion{N}{2}] $\lambda\lambda$6548,6584 lines,
indicative of a very low abundance.  Also note the detection of the [\ion{S}{3}] lines
at $\lambda\lambda$9069, 9532 and the emission lines from the Paschen sequence
which are labeled P14 - P8.
The noise spikes at wavelengths above $\lambda$8500 are due to residuals from subtraction
of the atmospheric emission lines.
H$\alpha$ is off scale to allow better visibility of the weaker emission lines.
}
\label{fig:fig3}
\end{figure}


Three 900 second exposures were taken at an airmass values less than $\sim$1.1, with the slit at a 
fixed position angle which approximated the parallactic angle at the midpoint of the observation.
Wavelength calibrations were derived from Hg(Ar), Ne, Xe, and Kr lamp spectra taken with the
telescope at zenith.  MODS is equipped with a closed-loop active
flexure compensation system, so only a small ($\lesssim$ 1\AA) residual
flexure correction, calculated from night-sky lines in the spectra,
was applied to the final wavelength calibration.

Standard procedures within the IRAF 2-D package (\texttt{twodspec}) were used to bias-subtract, 
flat-field, and illumination-correct the raw data frames.
Airmass dependent atmospheric extinction was corrected using the Kitt
Peak \citep{crawford70} and Cerro Paranal \citep{patat11}
extinction curves corrected approximately to the 3220-meter elevation
of Mt.\ Graham adopting an atmospheric scale height of 7\,km.

The multiple sub-exposures were combined, eliminating cosmic rays in the process.
The resulting images were then flux-calibrated using the sensitivity curve derived from
the standard star observation.
Finally, a trace fit to the continuum source in the slit was used to extract the
galaxy emission within an aperture that encompassed $\gtrsim99\%$ of the light.

In Figure~\ref{fig:fig2} we present the calibrated blue spectrum and in Figure~\ref{fig:fig3} the calibrated red spectrum.
Figure~\ref{fig:fig2} shows the clear detection of the Balmer emission line sequence down to H12 at $\lambda$3750 due
to the relatively high sensitivity of LBT/MODS at wavelengths below $\lambda$4000 (and continuing down to the
atmospheric cut-off) and the relatively high spectral resolution.  Note also the high significance of the temperature 
sensitive [\ion{O}{3}] $\lambda$4363 emission line, the detection of the \ion{He}{1} $\lambda$5015 line,
and the lack of obvious underlying absorption for the Balmer lines.
Figure~\ref{fig:fig3} shows the very weak strength of the [\ion{N}{2}] $\lambda\lambda$6548,6584 lines,
indicative of a very low abundance, detection of \ion{He}{1} $\lambda$6678 at high significance, and the
detection of the [\ion{S}{3}] lines at $\lambda\lambda$9060, 9532 with emission lines from the Paschen sequence.


\section{EMISSION FLUXES AND PHYSICAL CONDITIONS}\label{sec:lines}


\subsection{Emission Line Measurements}\label{sec:iraf}
Emission line strengths were measured from the two spectra using standard methods available within IRAF.
In particular, the \texttt{splot} routine was used to analyze the extracted one-dimensional spectra 
and to measure emission line fluxes by integrating under the line. In cases where emission lines were 
blended, multiple Gaussian profiles were fitted to derive the integrated line fluxes. 
Special attention was paid to the Balmer lines, which, in some extragalactic \ion{H}{2} regions, are located in 
troughs of significant underlying stellar absorption \citep[e.g.,][]{berg11}. 
For Leo~P there is very little evidence of underlying absorption.
For the bluest Balmer lines, multiple component fits were attempted in which the absorption was fit by a broad, 
negative Lorentzian profile and the emission was fit by a narrow, positive Gaussian profile.  
However, even in these cases, the correction for underlying absorption was small, i.e., comparable to the 
uncertainty in the flux in the line, so simple integration under the line was used.

The errors of the flux measurements were approximated using 
\begin{equation}
	\sigma_{\lambda} \approx \sqrt{ {(2\times \sqrt{N}\times \xi)}^2 + {(0.02\times F_{\lambda})}^2 } ,	\label{eq:uncertainty}
\end{equation}
where $N$ is the number of pixels spanning the Gaussian profile fit to the narrow emission lines
and $\xi$ is the rms noise in the continuum 
determined as the average of the rms on each side of an emission line. 
For weak lines, the uncertainty is dominated by error from the continuum subtraction, 
so the rms noise term determines the uncertainty.
For the lines with flux measurements much stronger than the rms noise of the continuum 
(usually the H$\alpha$ lines and often the [\ion{O}{3}] $\lambda\lambda$4959,5007 doublet) 
the error is dominated by flux calibration and de-reddening uncertainties. 
In this case, a minimum uncertainty of 2\% was assumed \citep[based on the uncertainties in the standard star
measurements,][]{oke90}, and the right hand term above dominates the uncertainty estimate.


\subsection{Reddening Corrections}\label{sec:redcor}
The relative intensities of the Balmer lines are used to solve for the reddening using the reddening 
law of \citet{cardelli89}, assuming $A_{V}=3.1\ E(B-V)$ and 
the theoretical case B values from \citet{storey87} interpolated to the temperatures
derived from the [\ion{O}{3}] lines. 
We used a minimized $\chi^2$ approach to solve simultaneously for the reddening and underlying absorption based on 
the H$\alpha$/H$\beta$, H$\gamma$/H$\beta$, and H$\delta$/H$\beta$ ratios \citep[cf.][]{os01}.
The reddenings derived for the two spectra are both very low with
C(H$\beta$) $\lesssim$ 0.1.
The underlying Balmer absorption is consistent with zero 
for both spectra.  
The reddening correction for the LBT/MODS spectrum can be tested by comparing the 
corrected higher numbered Balmer lines to their theoretical values.  
The H9 $\lambda$3935 and H10 $\lambda$3798 corrected fluxes are consistent with their theoretical
ratios to H$\beta$ of 0.074 and 0.054, respectively.
The derived values are listed in Table~\ref{tbl1}.  Note that in Section~\ref{sec:he} we will
re-derive the reddening (and other parameters) from the combination of H and He lines. 

The foreground reddening in this direction is estimated to be A$_V$ = 0.07 magnitudes \citep{schlafly11}, 
which is very weak, 
corresponding to the high Galactic latitude of Leo~P ($+$54$\arcdeg$).  The total reddening in the
LBT/MODS spectrum  
of C(H$\beta$) $=$ 0.09 corresponds to an A$_V$ $\approx$ 0.2. This is slightly larger than the  
expected foreground reddening, indicating only a small amount of reddening internal to the galaxy, 
in accordance with its very low metallicity \citep[e.g.,][found a value of A$_V$ = 0.13
magnitudes of internal extinction in I~Zw~18]{cannon02}.
The relative flux estimates and corresponding errors for the two spectra,
corrected for reddening, are listed in Table~\ref{tbl1}. 

Overall, there is good agreement between the relative fluxes in the emission lines.  
A priori, some small differences are expected due to the difference in slit widths and/or a 
slight difference in slit placement between the two observations.  
One small difference occurs for the fainter Balmer lines H11 ($\lambda$3771), H10 ($\lambda$3798), 
H9 ($\lambda$3825), H8$+$HeI ($\lambda$3889), and H7$+$[Ne~III]. 
For these five emission lines, the KPNO 4-m fluxes are systematically less (though not significantly) 
than those from the LBT/MODS spectrum.
We believe this to be an effect of the very weak underlying H absorption (since observed H absorption is nominally
constant in terms of equivalent width, the weakest, i.e., bluest, lines are the most sensitive to the 
presence of underlying absorption). Because the resolution of the KPNO 4-m spectrum is roughly half that of 
the LBT/MODS spectrum, the broader underlying absorption will have a greater impact on the H emission line fluxes. 
The overall consistency between the abundances derived from the two spectra in Section \ref{sec:abund} is a
direct result of the good agreement between the two spectra.


\subsection{Electron Temperature and Density Determinations}\label{sec:phys}

For the purpose of deriving nebular abundances, 
we adopt a simple two zone approximation, where $t_2$ and $t_3$ are the electron
temperatures (in units of $10^{4}$ K) in the low and high ionization zones respectively.
For the high ionization zone, the [\ion{O}{3}] I($\lambda\lambda$4959,5007)/I($\lambda$4363)
ratio was used to derive a temperature.
We use the 5-level atom calculations available with the IRAF task \texttt{temden}.
The derived temperatures are given in Table~\ref{tbl2}.
We derive relatively high temperatures of 17,150$^{+2040}_{-1390}$ K from the KPNO 4m spectrum
and 17,350 $^{+1390}_{-1060}$ K from the LBT/MODS spectrum, as expected for
a low metallicity \ion{H}{2} region.  The agreement between the two independent 
measurements indicates that the \ion{H}{2} region in Leo~P is clearly very low metallicity.

Because neither the [\ion{O}{2}] $\lambda\lambda$7320,7320 lines nor the 
[\ion{N}{2}] $\lambda$5755 line were detected at high significance, we cannot
derive a temperature for the low ionization zone directly and, therefore,
need to assume a temperature.
We used the relation between $t_{2}$ and $t_{3}$ proposed by \citet{pagel92},
based on the photoionization modeling of \citet{stasinska90} to determine
the low ionization zone temperature:
\begin{equation}
        {t_{2}}^{-1} = 0.5({t_{3}}^{-1} + 0.8).
	\end{equation}
The [\ion{S}{3}] $\lambda$6312 emission line was only detected at the 2 $\sigma$ level, 
so for the temperature in the [\ion{S}{3}] zone we assumed a temperature based on
the relationship derived by \citet{garnett92}:
\begin{equation}
        t(S^{+2}) = 0.83(t_{3}) + 0.17.
	\end{equation}
The low and high ionization region temperatures are tabulated in Table~\ref{tbl2}.

[\ion{S}{2}] $\lambda\lambda$6717,6731 and [\ion{O}{2}] $\lambda\lambda$3726,3729
were used to determine electron densities.  These densities are both
consistent with the low density limit.  The [\ion{O}{2}] measurement, which is allowed 
by the higher resolution of the LBT/MODS spectrograph, provides a stronger constraint
on the density due to the lower critical densities of the [\ion{O}{2}] emission lines \citep{of06}. 
For all abundance calculations we assume $n_{\rm e}=10^2$ cm$^{-3}$ (which is consistent
with the 1 $\sigma$ upper bounds and produces identical results for all lower values of $n_{\rm e}$).

\section{NEBULAR ABUNDANCE ANALYSIS}\label{sec:abund}


We analyze both spectra separately, compare the derived abundances, and provide adopted 
abundances and uncertainties in Table~\ref{tbl2}.
Calculated errors in the abundances provide a statistical estimate only.
Additional errors may be important, such as systematic errors due to temperature 
fluctuations \citep[cf.,][]{pena12}.
However, adopting this approach means that the derived abundances and uncertainties will be 
directly comparable to most of those reported in the literature.

\subsection{Oxygen Abundance Determination}\label{sec:oxygen}

We determine oxygen abundances based on our estimated two zone electron temperatures.  
Ionic abundances were calculated with:
\begin{equation}
	{\frac{N(X^{i})}{N(H^{+})}\ } = {\frac{I_{\lambda(i)}}{I_{H\beta}}\ } {\frac{j_{H\beta}}{j_{\lambda(i)}}\ }.
	\label{eq:Nfrac}
\end{equation}
The emissivity coefficients, which are functions of both temperature and density, were determined using 
the IONIC routine in IRAF with atomic data updated as reported in \citet{bresolin09}.
This routine applies the 5-level atom approximation, assuming the appropriate 
ionization zone electron temperature.
The total oxygen abundance, O/H, is the sum of O$^+$/H$^+$ and O$^{++}$/H$^+$.

The oxygen abundance determinations for Leo~P are given in Table~\ref{tbl2}. 
We derive an oxygen abundance of 12 + log(O/H) = 7.17 $\pm$ 0.07 from the KPNO 4-m
spectrum and 7.17 $\pm$ 0.05 from the LBT/MODS spectrum.  The results from the two
spectra are in good agreement, and we adopt the error weighted average of 12 + log(O/H) = 7.17 $\pm$ 0.04
as the measurement of the oxygen abundance.

The oxygen abundance in Leo~P is one of the lowest oxygen abundances ever derived for an \ion{H}{2} region.
Formally, the oxygen abundance in Leo~P is equivalent to that of the famous low metallicity
galaxy I~Zw~18 \citep[with 12 + log(O/H) = 7.17 $\pm$ 0.04;][]{skillman93, izotov99} and
lower than that of SBS~0335-052E \citep[log(O/H) = 7.33 $\pm$ 0.01;][]{izotov97}.  It is slightly 
higher than observed in SBS~0335-052W \citep[log(O/H) = 7.12 $\pm$ 0.03;][]{izotov05}
and DDO~68 \citep[log(O/H) = 7.14 $\pm$ 0.03;][]{pustilnik05, izotov07b}, which are the current 
record holders.  

Note that there are some reports of even lower oxygen abundances in the literature, but
these are typically from spectra with lower signal/noise ratios.  A small overestimate of the [\ion{O}{3}]
$\lambda$4363 line can produce artificially high values of the electron temperature, which
translate into lower oxygen abundances.  These objects are usually associated with 
derived electron temperatures well in excess of 20,000 K, which is difficult to achieve 
in real \ion{H}{2} regions, even with only trace amounts of oxygen present.

The significance of the low oxygen abundance in Leo~P will be discussed further in Section~\ref{sec:diss}.

\subsection{Nitrogen Abundance Determination}\label{sec:nitrogen}

We derive the N/O abundance ratio from the [\ion{O}{2}]$\lambda$3727/[\ion{N}{2}]$\lambda$6584 ratio and
assume N/O = N$^{+}$/O$^{+}$ \citep{peimbert69}.
\cite{nava06} have investigated the validity of this assumption.  
They concluded that although it could be improved upon with modern photoionization models, 
it is valid to within about 10\%.
Thus, we employ this assumption, mostly for the purposes of direct comparison with other 
studies in the literature.

The nitrogen to oxygen relative abundance determinations are given in Table~\ref{tbl2}.
For Leo~P, log(N/O) $=$ $-$1.40 $\pm$ 0.09 from the KPNO 4-m spectrum and 
$-$1.33$\pm$0.05 from the LBT/MODS spectrum.  Again, there is excellent agreement between the two
independent measurements. We adopt an error weighted mean of $-$1.36 $\pm$ 0.04.

The value of N/O in Leo~P is high relative to most determinations of N/O at very low O/H, and significantly
higher than the very narrow plateau at $-$1.60 $\pm$ 0.02 which \citet{izotov99} highlighted
in their study of very low metallicity emission line galaxies.  This result and its implications
for nitrogen nucleosynthesis will be discussed in Section~\ref{sec:no}.

\subsection{Neon, Sulfur, and Argon Abundances}

To estimate the neon abundance, we assume that Ne/O =  Ne$^{++}$/O$^{++}$
\citep{peimbert69}.  
The neon to oxygen relative abundance determinations are given in Table~\ref{tbl2}.
We derive log(Ne/O) = $-$0.72 $\pm$ 0.05 from the KPNO 4-m spectrum and $-$0.78 $\pm$ 0.04
from the LBT/MODS spectrum.  These determinations are nearly identical and
we adopt $-$0.76 $\pm$ 0.03 as our final determination.

To determine the sulfur and argon abundances, for direct comparison, we adopt the ionization 
correction factors (ICF) of \citet{thuan95} from \ion{H}{2} region photoionization models 
to correct for the unobserved S$^{+3}$, Ar$^{+2}$, and Ar$^{+4}$ states.
In the KPNO 4-m spectrum,  no [\ion{S}{3}] emission lines were detected, so we do not
calculate a sulfur abundance as the ICF becomes too uncertain \citep{garnett89}, and 
Ar$^{+3}$ $\lambda$7136 was beyond the usable wavelength coverage of the spectrum.  Thus,
we only have S/O and Ar/O relative abundance measurements from the LBT/MODS spectrum.
For log(S/O) we obtain $-$1.49 $\pm$ 0.07 and for log(Ar/O) we obtain $-$2.00 $\pm$ 0.09.

\subsection{The He Abundance of Leo~P from LBT/MODS}\label{sec:he}

The LBT/MODS spectrum of Leo~P shows the detection of several \ion{He}{1} emission lines
at high significance.  Thus, we can use the methodology developed in \citet{os04} and \citet{aos1, aos2} to
derive a helium abundance.  Specifically, we use a Markov Chain Monte Carlo (MCMC) method to 
efficiently explore the parameter space in physical conditions (i.e., temperature, density,
neutral hydrogen fraction) and other observable effects on the spectrum (i.e., reddening, absorption underlying
the H and He emission lines, optical depth in the He emission lines, collisional excitation of 
H and He emission lines).  We use the electron temperature derived from the [O III] emission lines 
as a prior, in a very conservative manner \citep[see discussion in][]{aos2}, producing negligible bias and effectively 
eliminating non-physical false minima.  In this way we determine the helium abundance, the 
physical parameters, and the uncertainties derived from observations of the nebula.

The LBT/MODS produces spectra with advantages over the spectra typically used to determine
nebular He abundances.  The vast majority of nebular He abundance determinations in the 
literature have been derived with a spectral resolution of roughly 6 - 8 \AA.  The higher
resolutions of the LBT/MODS spectrographs (2.4 and 3.4 \AA ) allow us to directly 
measure the absorption underlying the 
H emission lines and, because the resolution is a good match to the intrinsic width of
the emission lines, provide optimal sensitivity for the weak emission lines. 
Additionally, because LBT/MODS is a double spectrograph, all emission lines are observed
simultaneously, unlike some spectrographs where one needs to make multiple observations
in order to cover the wavelength range at the appropriate spectral resolution. 
Table~\ref{tbl3} presents the fluxes, EWs, and uncertainties for the H and He emission
lines used in this analysis.

The methodology developed in \citet{aos1, aos2} made use of helium emissivities calculated by
\cite{porter05, porter07}.  Thus, in all regards, we follow the
methodology of \citet{aos1, aos2}.  For example, H8 and \ion{He}{1} $\lambda$ 3889 are
deblended in a self-consistent way \citep{aos1} accounting for underlying absorption
and the equivalent widths of the stellar absorption underlying the \ion{He}{1} emission
lines are assumed to be equal \citep{aos1}.
Recently, \cite{porter12} have produced a new set of helium 
emissivities, which, in principle, represent an improvement on the older emissivities.
Unfortunately, an error has been discovered in the new emissivities \citep{porter13}.
We are currently working on a comprehensive paper to determine the effect of these corrected
new emissivities on helium abundance calculations, and an updated analysis of Leo~P
will appear in that paper.  
By using the \cite{porter05, porter07} emissivities in the present analysis, we can compare 
the derived He abundance directly to those in the literature in Section~\ref{sec:Yp}. 

Table~\ref{tbl4} shows the results of two different helium abundance determinations.  The
first solution in column 2 is the ``standard'' analysis, following \citet{aos2, aos3}, 
based on 4 \ion{H}{1} emission lines 
(H$\delta$, H$\gamma$, H$\beta$, and H$\alpha$) and 6 \ion{He}{1} emission lines
($\lambda\lambda$ 3889, 4026, 4471, 5876, 6678, 7065).  (Note that in Table~\ref{tbl4}, this is
referred to as nine emission lines because all are referenced to H$\beta$.) 
In column 3, we give the 
solution after adding in \ion{He}{1} $\lambda$5015.  The $\lambda$5015 emission line has 
typically not been used because of its close proximity to [\ion{O}{3}] $\lambda$5007 
results in a blend at lower resolution.  Note, however, that the resolution of the
SDSS spectra allowed \citet{izotov07a} to measure the \ion{He}{1} $\lambda$5015 in
their very large sample of low metallicity \ion{H}{2} regions.

Table~\ref{tbl4} shows stable, well-constrained solutions for the physical parameters and the He abundance.
As expected, and consistent with previous He abundance analyses, the derived electron temperature of 
17,100 K is in excellent agreement with the electron temperature derived from the [\ion{O}{3}] emission lines
(17,350 K), and significantly higher than the estimated low ionization zone temperature (14,530 K).  
The electron density is in the low density limit consistent with the [\ion{S}{2}] and [\ion{O}{2}] emission lines.  
The reddening, C(H$\beta$) is in excellent agreement with that derived from the H lines alone (for
this analysis we use the raw fluxes, not corrected for reddening, so the reddening determinations
are independent).
The solution for the underlying H absorption is in excellent agreement with the number previously
derived from the H lines alone.
Because the favored value for the optical depth in the He lines is low, the 
solution is free from assumptions about how the optical depth effects are modeled.  
Also, the favored solution shows little evidence for neutral H, so the effects of 
collisional excitation of the lower Balmer lines are also negligible.
The $\chi$$^2$ value of 3.3 (confidence level of 93\%)  for the ``standard'' analysis is well 
within the cut-off of 4.0 per degree of freedom (confidence level of 94.5\%) for goodness of fit 
recommended by \citet{aos3}.

With the addition of the $\lambda$5015 \ion{He}{1} emission line,
the solution for the He abundance is essentially unchanged 
(as might be expected since the line has an observational
uncertainty of 20\%).  However, there is a small decrease in the EW of the underlying He 
absorption. The addition of $\lambda$5015 is expected to have the greatest impact on the EW of the
underlying He absorption because it is an intrinsically weak \ion{He}{1} line. 
Thus, like $\lambda$4026, it is very sensitive to underlying absorption; however, because $\lambda$5015 is a singlet line 
(unlike $\lambda$4026, which is a triplet line), it is not sensitive to radiative transfer absorption. 
Note that \citet{izotov07a} have compared the $\lambda$5015 emission line strength to
$\lambda$6678 in their very large sample of low metallicity \ion{H}{2} regions, 
and find generally lower ratios of $\lambda$5015/$\lambda$6678 when compared to the 
theoretically expected values.  They interpret this as a
departure from the standard assumption of case B (resonance lines are optically thick).
For this quality of spectrum, (without detection of \ion{He}{1} $\lambda$3864) 
it is not possible to further test the possibility of a departure from the 
case B assumption. 
Our detection of $\lambda$7281, though not strong, is in good agreement with the 
case B theoretical prediction.
Note that if $\lambda$5015 were weaker due to
the departure from case B, that we would expect an {\it increase} in the detected
underlying absorption as opposed to the slight decrease which results.
Thus, we choose to present an analysis including $\lambda$5015.
Adding $\lambda$5015 results in a slightly higher $\chi$$^2$ of 3.5, 
and a significantly lower $\chi$$^2$ per degree of freedom of 1.75
(corresponding to a 83\% confidence level). 

Two possible concerns for converting a He$^+$/H$^+$ ratio into a He/H ratio are the
presence of He$^{++}$ or the presence of neutral He.  The \ion{He}{2} $\lambda$4686 emission line 
is not detected in our spectra at high confidence, indicating that any He$^{++}$ contribution is
negligible.  \citet{pagel92} promoted the use of the ``radiation softness parameter'',
$\eta$ $\equiv$ (O$^+$/S$^+$)(S$^{++}$/O$^{++}$) \citep{vp88}, as an indication whether
a correction for neutral He is warranted.  Based on photoionization models with
model stellar atmospheres, \citet{pagel92} concluded that for values of log($\eta$) $<$ 0.9
the correction of neutral He was negligible.  From the LBT/MODS spectrum, $\eta$ $=$
1.84 $\pm$ 0.64, which corresponds to log($\eta$) $=$ 0.26, well below the region
where one needs to make corrections for neutral He.  Thus, we will adopt He/H $=$ He$^+$/H$^+$.
The value of He$^+$/H$^+$ $=$ 0.0837$^{+0.0084}_{-0.0054}$  from the last column of Table~\ref{tbl4} converts 
to a value of 0.2509$^{+0.0184}_{-0.0123}$  for the He mass fraction.


\begin{figure}[h]
\plotone{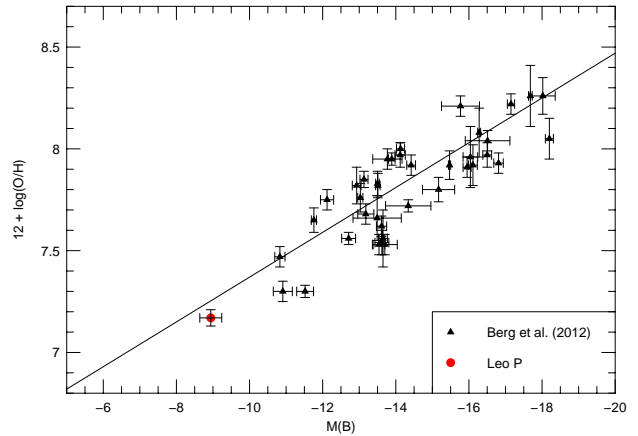}
\caption{The newly derived oxygen abundance and estimated absolute magnitude for Leo~P
\citep[from][estimating a distance of 1.75 $\pm$ 0.25 Mpc]{rhode12}  compared to the B-band luminosity/oxygen 
abundance relationship for dwarf star forming galaxies from \citet{berg12}.  
For the Berg et al.\  sample, only galaxies with reliable distances (e.g., TRGB distances) 
and direct abundances were included (i.e., the ``Combined Select'' sample).  The solid line is the 
regression from \citet{berg12}.
}
\label{fig:fig4}
\end{figure}


\section{THE SIGNIFICANCE OF THE CHEMICAL ABUNDANCES IN LEO~P}\label{sec:results}

\subsection{The Luminosity-Metallicity (L-Z) Relationship and Leo~P}\label{sec:lz}

Although only a single \ion{H}{2} region is observed in Leo~P, we will assume that the derived abundances
are representative of the ISM in the galaxy.  This follows from the observation that
metallicity gradients and variations are observed to be small or non-existent in
low-mass galaxies \citep[e.g.,][]{skillman89, ks96, ks97, lee06b, croxall09, berg12}.

\citet{berg12} have compiled a sample of nearby dwarf galaxies with accurate distances measured
from either the tip of the red giant branch (TRGB) or Cepheid variable stars and oxygen abundances
measured from the ``direct'' method (their ``Combined Select'' sample).  
They have shown that there is relatively low scatter in
the relationships between the oxygen abundance with the B-band luminosities, the 4.5 $\mu$m luminosities,
and the masses inferred from infrared luminosity and color.  The L-Z relationship is best
understood as a fundamental relationship between galaxy mass and chemical abundance \citep{tremonti04}.
The low abundances in dwarf galaxies reflect inefficient chemical evolution due to large
gas mass fractions and the reduced capacity to retain newly synthesized heavy elements.

In Figure~\ref{fig:fig4}, using the
estimated B-band luminosity of Leo~P from \citet{rhode12}, we compare the position of Leo~P with the
B-band luminosity/oxygen abundance relationship determined from the ``Combined Select'' sample by \citet{berg12}.  
The uncertainty in the B-band luminosity of Leo~P is dominated by the uncertainty in its distance.  
\citet{rhode12} have estimated
a distance of roughly between 1.5 and 2.0 Mpc from observations of the resolved stars.  This is supported by 
the distance estimate of 1.3$^{+0.9}_{-0.5}$ Mpc derived by \citet{giovanelli12} from the baryonic 
Tully-Fisher relationship. Adopting a distance estimate of 1.75 $\pm$ 0.25 Mpc and the photometry from
\citet{rhode12} results in a B-band luminosity of $-$8.94 $\pm$ 0.30 for Leo~P.

Figure~\ref{fig:fig4} shows the excellent agreement between the measurements for Leo~P and the L-Z 
relationship of \citet{berg12}.  
In this regard,
Leo~P represents an extension of this relationship to very low luminosities for actively star
forming dwarf galaxies.
It would be very valuable to have a more secure distance to Leo~P in order to confirm this result.
Recent ground-based imaging observations (K. McQuinn, private communication) have
been obtained that reach substantially deeper than the WIYN images from \citet{rhode12}. The RGB is definitively
detected in these images, and a preliminary
analysis of these new data produce a distance determination that is consistent with the range
quoted above. A thorough analysis and presentation of these new observations will be forthcoming.
In fact, future observations of distances to other low metallicity objects would be equally
valuable\footnote{For example, DDO~68 (also known as UGC~5340), with an estimated distance of $\sim$ 6 Mpc 
\citep[from brightest stars,][]{makarova98} and 10 Mpc \citep[after correcting the velocity for the
Local Void,][]{pustilnik11a}, does not yet have a published TRGB distance.
Note that Hubble Space Telescope observations have been acquired (program HST-GO-11578) and
a preliminary distance of 12.1 $\pm$ 0.7 Mpc has been derived (A.\ Aloisi, private communication).}.

The \citet{berg12} sample, based on the nearest galaxies, is essentially a volume limited sample.
Thus, the relationships from this sample should be representative of typical galaxies (with the 
exception of galaxies experiencing severe environmental influences).  The implication is that the ISM oxygen abundance
of a star forming galaxy is a very strong function of the stellar mass, down to the least
massive star forming galaxies known.  Given that the luminosity function for galaxies predicts large numbers
of these very low mass galaxies, very low metallicity galaxies like Leo~P should be quite common.  
However, the requirement that they possess relatively high surface brightness \ion{H}{2} regions
for abundance analysis may limit their detectability significantly.  


\subsection{The $\alpha$ Abundances and Leo~P}\label{sec:alpha}

\begin{figure}[ht!]
\epsscale{1.0}
\plotone{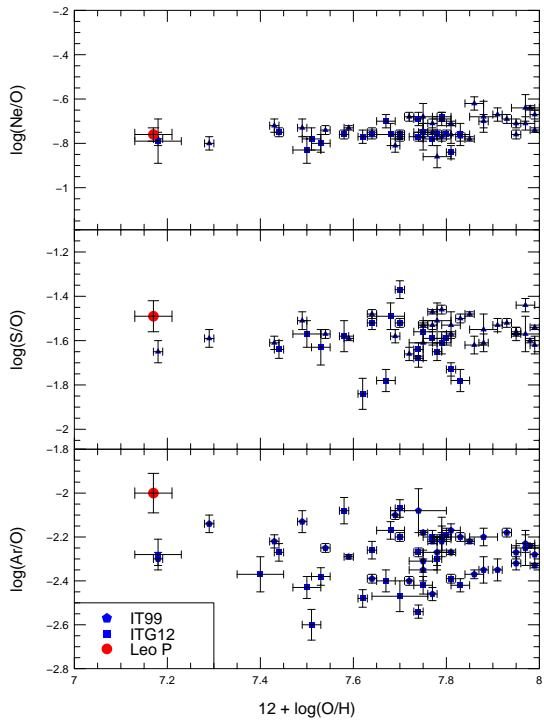}
\caption{The Ne/O, S/O, and Ar/O ratios in Leo~P compared to the observed ratios in emission line galaxies from
the samples of \citet{izotov99} and \citet{itg12}.  Data points with uncertainties of greater than 0.1 dex have been
excluded for clarity.  Note that all three panels have a y-axis range of 1 dex.
}
\label{fig:fig5}
\end{figure}

Figure~\ref{fig:fig5} shows the Ne/O, S/O, and Ar/O relative abundances in Leo~P as compared with
the relative abundances in emission line galaxies as studied by \citet{izotov99}. 
For clarity, data points with uncertainties of greater than 0.1 dex have been excluded.
Because $\alpha$
element production is thought to be dominated by core collapse supernova production, under the assumption of 
a universal mass function (IMF), the ratios of the alpha elements are expected to be constant
as a function of metallicity.  The Ne/O ratio in Leo~P is consistent with the mean value and scatter of the very
metal poor emission line galaxies shown in Figure~\ref{fig:fig5}.  The S/O ratio in Leo~P is consistent 
with the scatter seen in the metal poor emission line galaxies.  The Ar/O for Leo~P appears a bit higher
than the typical emission line galaxies, although consistent with the scatter. Given that the ionization
correction for Ar/O \citep{izotov99} is the most uncertain (the errorbars in the observations do not 
account for uncertainties in the ionization corrections), 
at this time we cannot conclude that there is strong evidence for an anomalous Ar/O ratio in Leo~P. 
Note that \citet{stevenson93} have suggested that comparing the [\ion{Ar}{3}] $\lambda$7136 emission line to the
[\ion{S}{3}] $\lambda$9069 emission line as a measurement of Ar/S may be a more reliable method for testing
the stability of relative Ar abundances.  Our Ar/S measurement for Leo~P of $\sim$0.3 is elevated relative to the typical
value of 0.2 \citep{stevenson93, izotov99}, as indicated by Figure~\ref{fig:fig5}.

It is well established that both emission line dwarf galaxies and relatively quiescent 
(or low star formation rate) dwarf
galaxies show relatively constant values of Ne/O, S/O, and Ar/O \citep{thuan95, vanzee97a, izotov99,
vanzee06a}.
This has generally been regarded as expected under the assumption of a universal IMF; however, 
recent work revealing a trend of lower H$\alpha$-to-UV flux ratio with decreasing galaxy luminosity  
has brought that assumption into question for dwarf galaxies \citep{hg08, meurer09, jlee09, boselli09}.
The constancy of the elemental abundance ratios has not played a significant role in this 
debate, but clearly the observed trends favor a universal IMF. Thus, alternative explanations
of the H$\alpha$-to-UV flux ratio trend assuming a universal IMF \citep[e.g.,][]{fumagalli11, weisz12} 
are likely to be favored, in agreement with the general conclusion of \citet{bastian10} that there is 
little secure evidence supporting a non-universal IMF.



\begin{figure}[t]
\epsscale{1.0}
\plotone{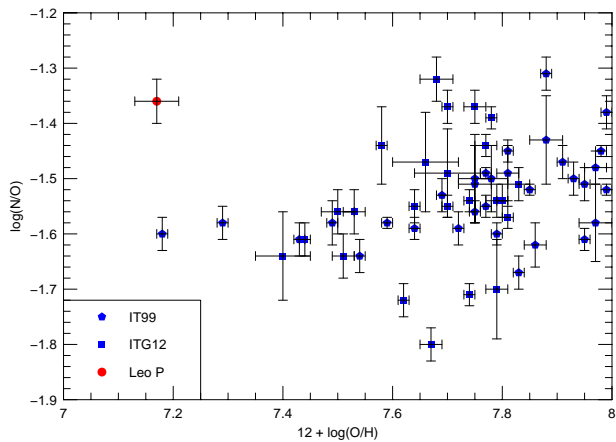}
\caption{The N/O ratio in Leo~P compared to the observed ratios in emission line galaxies from
the samples of \citet{izotov99} and \citet{itg12}.  Note that Leo~P lies significantly above the
very narrow plateau at log(N/O) $\approx$ $-$1.6 for the emission line galaxies with
12 + log(O/H) $<$ 7.6 previously noted by \citet{izotov99}.  This indicates the likely presence of
a secondary component of N in Leo~P.
}
\label{fig:fig6}
\end{figure}


\subsection{N/O Relative Abundances and Leo~P}
\label{sec:no}

\citet{garnett90} first showed that the N/O ratio in low metallicity star forming galaxies is relatively constant
as a function of O/H (with a mean value of log(N/O) = $-$1.46$^{+0.10}_{-0.13}$ for these ``plateau'' objects).
Although the mean is well defined, \citet{garnett90} also pointed out that the scatter in N/O at a given
O/H is larger than can be accounted for in terms of observational errors.  He suggested that the relatively
large scatter in N/O at low O/H could be understood in terms of the different delivery times of N and O. 
That is, O production is understood to be dominated by primary nucleosynthesis from massive stars and 
delivered early after a star formation event;  N can be produced by massive
stars (and delivered early with the primary oxygen) and by intermediate mass stars (and 
delivered relatively later).  N can have both primary and secondary origins. 
Thus, in a galaxy where the N/O ratio has been elevated from its initial N/O ratio from massive stars 
by contributions from delayed nitrogen from intermediate mass stars, a burst of star formation will
drive a galaxy to higher values of O and lower values of N/O, but later contributions from intermediate
mass stars will raise the N/O at a constant O/H.  
At higher metallicities (i.e., 12 + log(O/H) $\ge$ 8.0), the average N/O rises with increasing
O/H, and this is thought to be indicative of the increasing influence of secondary nitrogen production \citep{pagel85}.
This hypothesis has been supported quantitatively by simple models of galaxy chemical evolution by, e.g., 
\citet{matteucci85}, \citet{pilyugin93}, and \citet{fields98}.

In a subsequent study of  blue compact dwarf galaxies, \citet{izotov99} drew attention to a different plateau 
with a very small dispersion (0.02 dex) in log (N/O) (with a central value of $-$1.60) in the extremely 
metal-poor galaxies with 12 + log(O/H) $\le$ 7.6.
They proposed that the absence of time-delayed secondary production of N (and C) is consistent with 
the scenario that extremely metal-poor galaxies are now undergoing their first burst of star formation,
and that they are therefore young, with ages not exceeding 40 Myr.
They further asserted that this countered the commonly held belief that C and N
are produced by intermediate-mass stars at very low metallicities (as these stars would not have yet completed
their evolution in these lowest metallicity galaxies).

Later, \citet{vanzee06a} derived an average of log(N/O) = $-1.41$ for a sample of isolated dwarf irregular galaxies.
Their sample included extremely metal-poor objects, but there was no evidence for the very narrow plateau in N/O
at 12 + log(O/H) $\le$ 7.6 observed in blue compact galaxies by \citet{izotov99}. Interestingly, all of the extremely 
metal poor \ion{H}{2} regions had N/O values in agreement with the average value of log(N/O) = $-1.41$.
\cite{nava06} revisited the observed N/O plateau with a large set of objects and determined a mean value for
the N/O plateau of log(N/O) = $-$1.43 with a standard deviation of $^{+0.071}_{-0.084}$.
They further concluded from a $\chi^2$ analysis that only a small fraction of the observed scatter in N/O is intrinsic.
Based on a Monte Carlo analysis of the scatter in the N/O versus O/H diagram, \citet{henry06} 
concluded that one could not distinguish between a delayed nitrogen hypothesis or the hypothesis that 
nitrogen is produced by massive stars alone at low metallicity.  They did conclude that allowing galaxy ages below
250 Myr could not explain the plateau morphology.

Figure~\ref{fig:fig6} shows the N/O relative abundance in Leo~P as compared with
the N/O relative abundances in emission line galaxies as studied by \citet{izotov99} and a new sample
available from \citet{itg12}.
For clarity, the emission line galaxy samples have been trimmed of objects with uncertainties in log(N/O) of
more than 0.1 dex.  
Clearly, Leo~P lies well above the very narrow plateau identified in extremely metal poor emission line 
galaxies by \citet{izotov99}.  


\begin{figure}[h!]
\epsscale{1.0}
\plotone{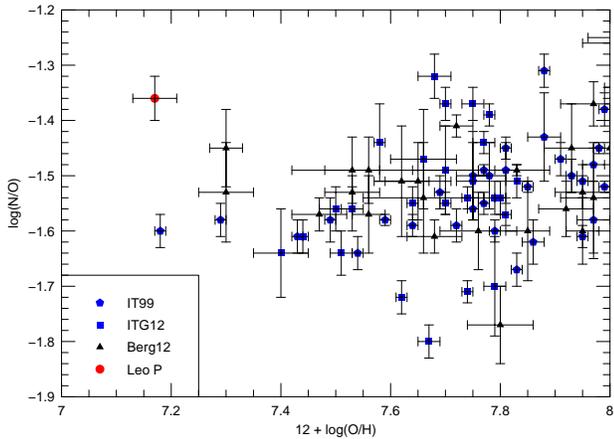}
\caption{The N/O ratio in Leo~P compared to the observed ratios in emission line galaxies from
the samples of \citet{izotov99} and \citet{itg12} and the nearby dwarf galaxies from the sample of \citet{berg12}.
Note the larger spread in log(N/O) at low values of log(O/H) for the nearby galaxies sample and that
the scatter in log(N/O) at higher values of log(O/H) for the nearby galaxies is comparable to that
of the emission line galaxies.
}
\label{fig:fig7}
\end{figure}


On the other hand, the N/O abundance for Leo~P lies within the scatter observed by \citet{garnett90} and 
\citet{vanzee06a}.
In Figure~\ref{fig:fig7}, we have added the galaxies from the sample of \citet{berg12} (again limited to those
with uncertainties in log(N/O) less than 0.1 dex).  In this diagram, the evidence for a very narrow plateau at values
of 12 + log(O/H) $\le$ 7.6 is not seen.  Thus, the very narrow plateau seen in the emission line galaxies 
cannot be simply a function of metallicity.
\citet{vanzee06b} and \citet{berg12} point out that there are nearby galaxies with values of 12 + log(O/H) $\le$ 7.6
and detailed star formation histories (derived from \textit{Hubble Space Telescope} observations of their 
resolved stars) that clearly show that the bulk of their star formation occurred well before
the last 40 Myr \citep[i.e., Leo~A and GR~8,][]{tolstoy98, cole07, dohmpalmer98, weisz11}. 
This implies that the 
young galaxy hypothesis is not a valid explanation for the plateau in N/O at low metallicity.

With the addition of Leo~P, it now looks as though even the most metal poor galaxies show a range in N/O.
If the scatter in the log(N/O) vs 12 + log(O/H) diagram is due to the time delay between
producing oxygen and secondary nitrogen as proposed by \citet{garnett90}, then it could be that the N/O in Leo~P is 
offset to a higher value 
due to a long period of relative quiescence in which the nitrogen
production is allowed to essentially complete before the next round of oxygen production.
That two of the three very low metallicity nearby galaxies in Figure~\ref{fig:fig7} 
lie above the very narrow plateau is interesting.
Note that under the interpretation that the scatter is due to secondary nitrogen production, then
the {\it lowest} values of N/O (not the mean value) represent the ratio of N/O in primary nucleosynthesis
by massive stars.

\citet{vanzee06a} looked at several variables for their possible influence on N/O abundance.
In particular, they found a correlation between N/O and color, in the sense that redder 
galaxies have higher N/O.
This supports the delayed release hypothesis of \citet{garnett90}, as currently quiescent
galaxies would be expected to have, on average, higher values of N/O.
\citet{berg12} recovered this trend in their sample.  For the B-V color of 0.36 for
Leo~P \citep{rhode12}, the relationship between B-V color and N/O determined by
\citet{berg12} predicts a value of log(N/O) = $-$1.50.  Although, formally, this is not consistent 
with our value of $-$1.36 $\pm$ 0.04, Leo~P falls well within the scatter in the relationship,
which has an intrinsic dispersion of 0.14 dex, and thus, follows the trend.

The question raised by the N/O versus O/H diagnostic diagram is whether emission line galaxies
and relatively quiescent dwarf galaxies occupy different sections of this diagram
and therefore represent different chemical evolution paths.  Within the literature there 
are proponents of the view that emission line galaxies (or blue compact galaxies) are intrinsically
different from relatively quiescent galaxies, and there are others who allow for the interpretation
that emission line galaxies represent a phase that many, or perhaps all, dwarf galaxies can 
pass through.  In the former interpretation, the comparison of relatively quiescent dwarfs with
emission line galaxies could be viewed as comparing two different families of galaxies.  
The emission line galaxies could represent a more homogeneous sample due to the fully 
populated IMFs normally associated with massive star formation events.  For the
latter interpretation, all of the points in the diagnostic diagram are directly comparable,
although the lower SFRs of relatively quiescent dwarfs are often associated with 
under-abundances of massive stars due to stochastic effects.
Unfortunately there are few high-quality spectra available for extremely low-metallicity galaxies.
Adding new observations remains a worthwhile enterprise.
The cause of the very narrow plateau in N/O in very low metallicity emission line galaxies 
is still lacking a definitive explanation, and
we will return to this in Section~\ref{sec:infall}. 


\begin{figure}[h!]
\epsscale{1.0}
\plotone{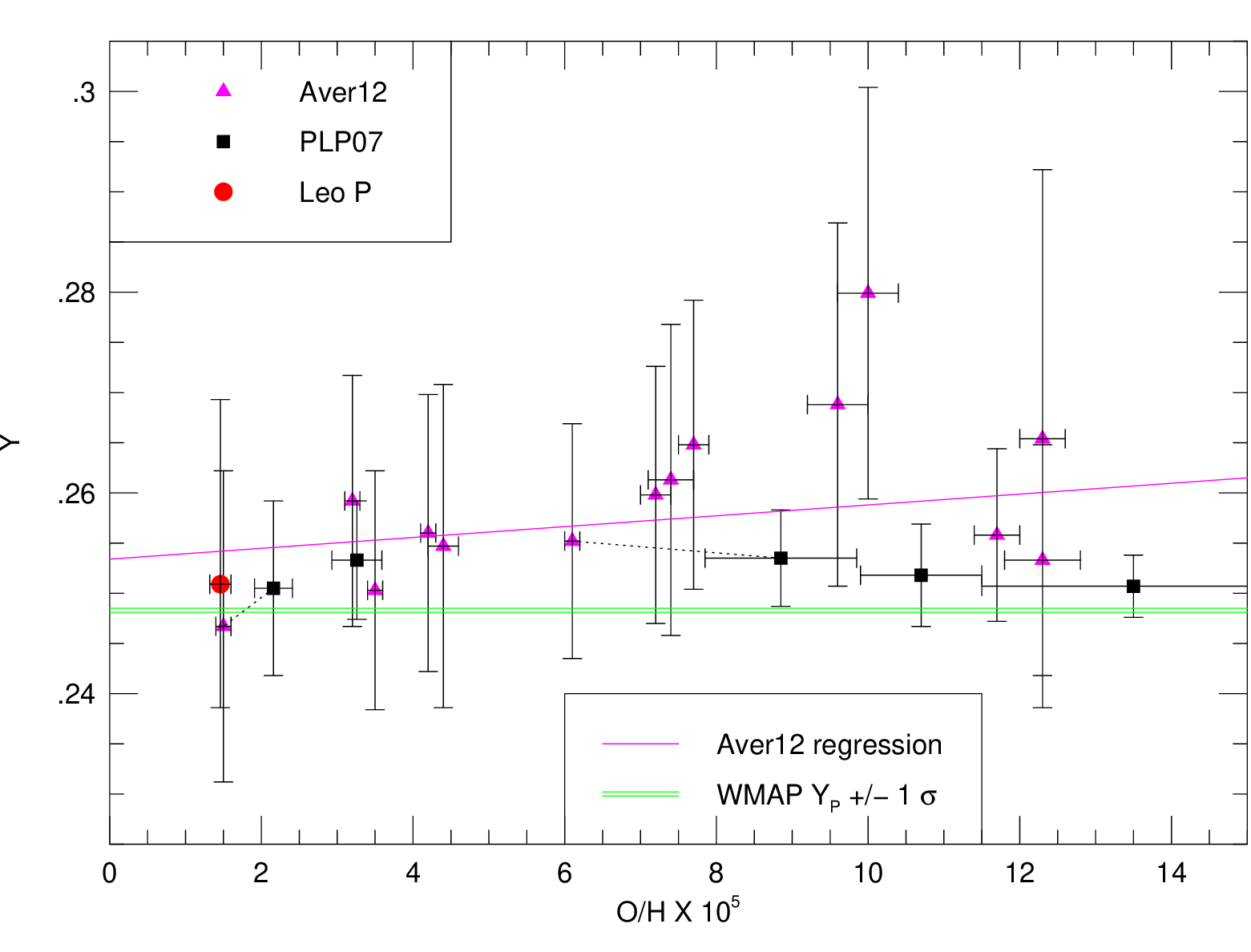}
\caption{The helium mass fraction (Y) and oxygen abundance for Leo~P compared to the abundances in emission line galaxies
from the sample of \citet{izotov07a} as analyzed by \citet[][Aver12]{aos3} and the sample from \citet[][PLP07]{plp07}.
The single line is the regression to the data from \citet{aos3}.
Note that the low value of O/H and the comparable error in Y for Leo~P make
this an important contribution in the determination of the primordial He abundance.
Two of the galaxies are common to both comparison samples, I~Zw~18 (at low O/H) and Haro~29 (also known as
I~Zw~36 and Mrk~209, at intermediate O/H) and their points are connected by dotted lines.
The narrow band marked WMAP7 is the range of values ($\pm$ 1$\sigma$) estimated for the primordial helium abundance following the calculation
by \citet{cyburt08} from the 7-year WMAP value for the baryon-to-photon ratio \citep{wmap11} and assuming the neutron mean
life from the Particle Data Group collaboration \citep{nakamura10}.
}
\label{fig:fig8}
\end{figure}


\subsection{Leo~P and the Primordial Helium Abundance}
\label{sec:Yp}

Next to the cosmic microwave background radiation, standard big bang nucleosynthesis (SBBN)
is the most robust probe of the early universe available  \citep[e.g.,][]{wssok}.
Furthermore, using the precise baryon density as determined by WMAP \citep{wmap11},
SBBN has effectively become a parameter-free theory \citep{cfo2}. As such, one can
use SBBN to make relatively precise predictions of the initial light element abundances of
D, $^{3}$He, $^{4}$He, and $^{7}$Li.
Therefore, an observational determination of these abundances becomes a test
of the concordance between SBBN theory and the analyses of microwave background
anisotropies. To test these predictions, the observed abundances must be determined
with high precision.  Unfortunately, there is a logarithmic relationship
between the baryon to photon ratio, $\eta$, and the primordial helium abundance, Y$_{p}$.
Thus, any meaningful test of the theory requires a determination of Y$_{p}$ to an accuracy
of a few percent. The 7-year WMAP value for $\eta$ is $(6.16 \pm 0.15) \times 10^{-10}$ \citep{wmap11}.
Using the procedures of \citet{cyburt08}, and assuming a neutron mean life of $885.7 \pm 0.8$ s 
\citep{nakamura10}, this translates to $Y_p = 0.2483 \pm 0.0002$, a relative uncertainty of only 
0.08\%\footnote{Note that the Planck Collaboration has just published a new prediction of the
primordial Helium abundance of 0.24771 $\pm$ 0.00014, corresponding to an uncertainty of 
0.06\% \citep{ade13}.}.

To date, the favored independent method of determining Y$_{p}$ uses observations of low metallicity
H~II regions in dwarf galaxies.  By fitting the helium abundance versus metallicity,
one can extrapolate back to very low metallicity, corresponding to the primordial
helium abundance \citep{ptp74}.  The oxygen to hydrogen ratio, O/H, commonly serves
as a proxy for metallicity.  The difficulties in calculating an accurate and precise
measure of the primordial helium abundance are well established \citep{os01,os04,aos2,izotov07a,plp07}.

Figure~\ref{fig:fig8} shows the
helium mass fraction (Y) and oxygen abundance for Leo~P compared to the abundances in emission line galaxies
from the sample of \citet{izotov07a} as analyzed by \cite{aos3} and the sample of very high 
quality spectra analyzed by \citet{plp07}. 
The two comparison data sets should be directly comparable in that they use identical atomic data for
converting relative emission line fluxes into abundances \citep[e.g.,][]{porter05, porter07}. However, 
the methodologies for determining physical conditions vary somewhat.  For example, two of the galaxies 
are common to both comparison samples, I~Zw~18 and Haro~29 (also known as
I~Zw~36 or Mrk~209).  These two galaxies are connected by solid lines in Figure~\ref{fig:fig8},
and one can see that the values for O/H are significantly different (due primarily to assumptions
about the effects of temperature variations within the nebulae).  We are not advocating that the
different samples be combined, but are only conducting the comparison to show the potential impact
of Leo~P on determinations of Y$_{p}$.  

Note that the low value of O/H and uncertainties in He/H which are comparable to the other best studied
nebulae in the literature make Leo~P an important contribution in the determination of Y$_{p}$.
To demonstrate this, we calculated a regression of Y on O/H by adding our Leo~P He abundance determination
to the sample from \citet{aos3}.  This reduced the intercept (Y$_p$) by roughly 0.3\% from 0.2534 to 0.2527
and the error on the intercept by roughly 8\% from 0.0083 to 0.0076.  The resulting regression of
\begin{equation}
Y = 0.2527 \pm 0.0076 + 61 \pm 92 (O/H)
\end{equation}
has a $\chi^2$ of 2.9.

The value of Y$_{p}$ implied by the 7-year WMAP observations is indicated in Figure~\ref{fig:fig8}. 
Note, especially, that our determination of Y for Leo~P is
in excellent agreement with the two determinations for Y in I~Zw~18 and that all three are in
excellent agreement with the 7-year WMAP prediction for Y$_p$.
Recently, \citet{it10} reported a significant difference
between the value of Y$_p$ derived from observations of metal poor \ion{H}{2} regions and the
value calculated from the cosmic microwave background observations and suggested this
was evidence for non-standard Big Bang Nucleosynthesis.  From the present analysis of the most
metal poor \ion{H}{2} regions, we see no motivation for non-standard Big Bang Nucleosynthesis.

\section{LEO P AND XMD GALAXIES}\label{sec:diss}
\subsection{The Search for Extremely Low Metallicity Galaxies}
\label{sec:XMD}

In an overview of the most metal-poor galaxies, \citet{ko00} defined ``very
metal deficient'' galaxies as those with metallicities of ten percent of
the solar value or less.  At the time, the solar value for the oxygen abundance
was generally accepted to be 12 + log(O/H) $=$ 8.91.  In the last decade, there has
been a consensus that the solar oxygen abundance is lower than previously
thought \citep[e.g., 12 + log(O/H) $=$ 8.69 $\pm$ 0.05,][]{asplund09} and
recent papers have adopted the definition of an ``extremely metal deficient''
(XMD) galaxy as having 12 + log(O/H) $\le$ 7.65 \citep[e.g.,][]{kniazev03,
pustilnik07, kakazu07, ekta08, brown08, ekta10b}.

Also, in the last decade, there have been several programs with the aim of
discovering more XMD galaxies.  The majority of these programs have
concentrated on surveying emission line galaxies.  Examples include
\citet{kniazev00, ugryumov03, kniazev03, melbourne04, izotov06, papaderos08,
brown08, guseva11}.  
I~Zw~18 \citep{searle72} and SBS~0335-052 \citep{izotov90} represent prototypes
for XMD galaxies that can be found in emission line surveys.  
Recently, direct oxygen abundance measurements have 
been extended to higher redshifts \citep{hu09, xia12}.
It is interesting that while recent surveys have significantly enlarged the
number of metal-poor galaxies, only a few XMD galaxies with oxygen abundances
below $\sim$5\% of the solar value are known, and no galaxies with abundance
below 12 + log(O/H) $\approx$ 7 have been discovered in the Local Volume.

Despite the many surveys of emission line galaxies, the number of XMD galaxies
remains small.  Searching via emission line surveys provides an extremely low 
yield.  As reported in \citet{itg12}, of one million SDSS spectra, 13,000 emission line
objects have detectable [\ion{O}{3}] $\lambda$4363, and, of these, there are only
15 candidates with 12 $+$ log(O/H) $\le$ 7.35.

An alternate approach to finding low metallicity galaxies relies on the
fundamental L-Z relationship between galaxian stellar luminosity (or mass) and abundance
for low redshift dwarf galaxies as discussed in Section \ref{sec:lz}
\citep[e.g.,][]{peimbert70, lequeux79, skillman89, lee06a, berg12}.
Because of the galaxy luminosity function, XMD galaxies are very numerous.
However, in part because of their small sizes, XMD galaxies with luminous, high surface 
brightness star forming regions are rare.  The result is an extreme paucity of 
XMD galaxies in emission line surveys.
Many XMD galaxies have been found in surveys of nearby, low luminosity galaxies.
XMD galaxies such as Leo~A \citep{skillman89,
vanzee06b}, SagDIG \citep{skillman89b, saviane02}, UGCA~292 \citep{vanzee00},
DDO~68 \citep{pustilnik05}, and two XMD galaxies in the Lynx-Cancer void
\citep{pustilnik11} have been found in this way.  Spectroscopic follow-up
of H$\alpha$ surveys of low luminosity galaxies and galaxies discovered in
blind \ion{H}{1} surveys continue to be very promising in this regard
\citep[e.g.,][]{cannon11}.
Surveys like ALFALFA have discovered hundreds of objects whose properties may be
confirmed by subsequent follow-up observations to be consistent with those of nearby,
low-luminosity dwarf galaxies.  A dedicated program to measure the nebular abundances
of these galaxies may be very fruitful in discovering new XMD galaxies.
%

\begin{figure}[h!]
\epsscale{1.0}
\plotone{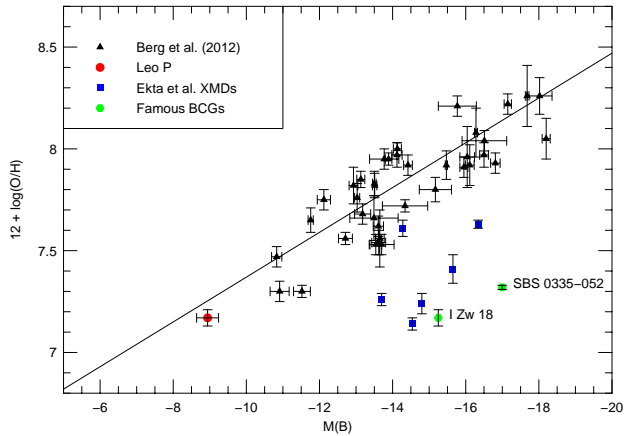}
\caption{Figure~4
is reproduced showing the positions of the famous XMD blue compact galaxies I~Zw~18 and SBS~0335-052 relative to the
B-band luminosity/oxygen abundance relationship for dwarf star forming galaxies from \citet{berg12}.
The XMD galaxies with disturbed kinematics as observed by \citet{ekta06, ekta08, ekta10a} are also shown.
Note the large offsets of the XMD emission line galaxies from the luminosity/oxygen abundance relationship defined by
dwarf star forming galaxies.
       }
\label{fig:fig9}
\end{figure}


\subsection{XMD Galaxies and the Infall of Metal-Poor Gas}
\label{sec:infall}

It is well-established that many of the XMD emission line
galaxies lie well off the L-Z relationship for dwarf galaxies.  This includes the
two best-known XMD galaxies, I~Zw~18 and SBS~0335-052.
In Figure~\ref{fig:fig9} we have reproduced the L-Z diagram shown in
Figure~\ref{fig:fig4}, and added I~Zw~18 and SBS~0335-052.  For I~Zw~18, we have used the 
new distance of \citet{aloisi07} and the abundances for the NW component from \citet{skillman93} which
is identical to that reported by \citet{izotovea99}. For SBS~0335-052, we have plotted the eastern (luminous) component
assuming a distance of 54 Mpc and used the oxygen abundance from \citet{izotovea99}.
In Figure~\ref{fig:fig9} it is clear that these two galaxies lie off the L-Z relationship
by about 7 magnitudes (or 0.7 dex in log(O/H)).  These offsets are much too large to be
attributable to increases in the luminosity due to the present starburst.

Based on new \ion{H}{1} observations of XMD galaxies, \citet{ekta10b} have proposed that 
inflow of metal-poor gas from the outskirts of a galaxy to the central star-forming regions 
(in the case of interaction), or cold gas accretion is the probable cause for the observed 
low emission-line metallicities of XMD galaxies.  This hypothesis is very appealing
because it solves two problems simultaneously.  It explains how small systems can achieve 
such high star formation rates that they become visible as emission line galaxies. It also
explains why so many of these galaxies do not follow the metallicity/luminosity 
relationship observed for normal dwarf galaxies in volume limited surveys.

Note that it is also possible to have very low abundances in galaxies through metal-enriched
outflows.  This is not as appealing as the \citet{ekta10b} hypothesis because this would 
require that the majority of nucleosynthesis in the XMD emission line galaxies is associated with
strong bursts of star formation resulting in galactic winds.  Furthermore, this does not
explain why other galaxies at similar masses have been able to enrich to higher metallicities.

In Figure~\ref{fig:fig9}, we have also added in the
XMD galaxies with disturbed kinematics as observed by \citet{ekta06, ekta08} and \citet{ekta10a}.  
Here one can see similarly large offsets of the XMD galaxies from the luminosity/oxygen abundance relationship 
defined by dwarf star forming galaxies.
Although offsets from the L-Z relationship for XMD emission line galaxies are well known, 
what Ekta and collaborators have added is the observation that all of these galaxies have
disturbed kinematics indicative of either interaction or infall. 
In support of this hypothesis, I~Zw~18 and SBS~0335-052 are both known to be associated
with disturbed \ion{H}{1} kinematics \citep{vanzee98, lelli12, pustilnik97, ekta09}. 

It would be very satisfying if this hypothesis could also explain the very narrow plateau seen in
the N/O abundances in XMD emission line galaxies.  One could hypothesize that if the XMD galaxies
are being polluted by relatively pristine gas, and if that pristine gas had a single value
of N/O, then XMD emission line galaxies would tend to have similar values of N/O.  
Unfortunately, absorption line studies of damped Lyman-$\alpha$ systems do not show a 
universal value of N/O; in fact they show a very large range in N/O at low values of
O/H \citep{pettini08} \citep[which, interestingly, has been used as evidence for a non-universal
IMF,][]{tsujimoto11}.  If all emission line XMD galaxies are the products of strong interactions
and infall, then it makes the very narrow plateau seen in the N/O abundances in XMD emission 
line galaxies even more difficult to understand.

The discovery of Leo~P supports the hypothesis that XMD galaxies are of two different types.
XMD emission line galaxies represent starbursts which are likely triggered by interaction with
a companion or infall from an \ion{H}{1} cloud, while relatively ``quiescent'' (i.e., galaxies with
low, but non-zero star formation rates) dwarf XMD galaxies are formed
through normal evolution of very low mass galaxies.


\section{CONCLUSIONS}
\label{sec:conclusion}

We have presented KPNO 4-m and LBT/MODS spectroscopic observations of an \ion{H}{2} region
in the nearby dwarf irregular galaxy Leo~P which was discovered recently in the Arecibo ALFALFA survey.
The results from both observations are in excellent agreement.
The KPNO and LBT observations yield accurate measurements of the temperature
sensitive [\ion{O}{3}] $\lambda$4363 line
and a ``direct'' oxygen abundance of 12 + log(O/H) = 7.17 $\pm$ 0.04.
This oxygen abundance is among the lowest ever measured in an \ion{H}{2} region.
Thus, Leo~P is an extremely metal deficient (XMD) galaxy.

For its estimated luminosity, Leo~P is consistent with the
relationship between luminosity and oxygen abundance seen in nearby dwarf galaxies.
Future observations of distances of nearby XMD galaxies will help to confirm that
the relationship between luminosity and oxygen abundance extends down to the 
least massive star forming galaxies known.

Leo~P shows normal $\alpha$ element abundances (Ne/O, S/O, and Ar/O) when compared to
other XMD galaxies.  These well-defined trends in the lowest metallicity galaxies
are supportive of the hypothesis of a universal initial mass function.

The N/O ratio in Leo~P is elevated relative to similarly low metallicity emission line
galaxies, but within the scatter seen in observations of large samples of dwarf galaxies.
The elevated N/O and optical color of Leo~P are consistent with the ``delayed release'' 
hypothesis for N/O abundances.

The high signal/noise ratio and spectral resolution of the LBT/MODS spectrum allow us
to derive a helium abundance with a precision of $\sim$ 5\% .  
We derive a helium mass fraction of 0.2509$^{+0.0184}_{-0.0123}$ 
which compares well with the WMAP + BBN prediction of 0.2483 $\pm$ 0.0002 for
the primordial helium abundance.  Supplementing the Leo~P observations with similar quality spectra
of other XMD galaxies will allow a future determination of the primordial 
helium abundance with an accuracy of less than two percent.

The discovery of Leo~P suggests that surveys of very low mass
galaxies compete well with emission line galaxy surveys for finding XMD
galaxies.  With present and future blind \ion{H}{1} surveys producing large numbers
of low mass galaxies, it should be possible to greatly increase the number of studies of 
abundances in XMD galaxies.

It is possible that XMD galaxies may be divided into two classes.  The XMD emission line
galaxies are rare.  This may be attributable to the uncommon occurrence
of starbursts triggered by infall or interaction.  Because of the galaxy luminosity 
function (which indicates large numbers of very low mass galaxies) and the relationship 
between galaxy stellar mass and oxygen abundance, 
XMD galaxies with very low chemical 
abundances due to their intrinsically small masses should be common.

Our ability to observe the XMD galaxies with intrinsically small masses 
may be limited by their low star formation rates and the
relatively short lives of the massive stars that produce the observable \ion{H}{2} regions.
Alternately, their intrinsically low luminosities may prevent them from inclusion in
most galaxy catalogs.  For example, the H$\alpha$ imaging survey of galaxies in the local 
11 Mpc volume by \citet{kennicutt08} had an apparent magnitude cut-off of $m_B$ $\le$ 15.
With an $m_B$ of 17.3 \citep{rhode12}, Leo~P is more than 2 magnitudes fainter than 
this cut-off.  Blind \ion{H}{1} studies may be the only efficient way to detect
these low luminosity XMD galaxies.


\acknowledgements
We are grateful to Fabio Bresolin for providing his updated atomic data
used in deriving the abundances.  We are also grateful for helpful input from
Antonio Peimbert, Dick Henry, and Simon Pustilnik.
We wish to thank the referee for a very helpful and prompt review which led to 
significant improvements in our paper.
EDS is grateful for partial support from the University of Minnesota and NSF Grant AST-1109066.
DAB is grateful for support from a Penrose Fellowship, a NASA Space Grant Fellowship, and 
a Dissertation Fellowship from the University of Minnesota.
RWP is supported in part by NSF Grant AST-1108693.
JMC is supported by NSF Grant AST-1211683.
KAO is supported in part by DOE grant DE-FG02-94ER-40823.
The ALFALFA team at Cornell is supported by NSF grants AST-0607007 and AST-1107390
to RG and MPH and by grants from the Brinson Foundation.
EAKA is supported by an NSF predoctoral fellowship.
KLR is supported by an NSF Faculty Early Career Development (CAREER) award (AST-0847109).

Observations reported here were obtained at the LBT Observatory, a joint
facility of the Smithsonian Institution and the University of Arizona.
LBT observations were obtained as part of the University of Minnesota's 
guaranteed time on Steward Observatory facilities through membership in 
the Research Corporation and its support for the Large Binocular Telescope,
and granted by NOAO, through the Telescope System Instrumentation Program (TSIP).
TSIP is funded by the National Science Foundation.
This paper uses data taken with the MODS spectrographs built with funding from NSF grant AST-9987045 and the 
NSF Telescope System Instrumentation Program (TSIP), with additional funds from the Ohio Board of Regents 
and the Ohio State University Office of Research.

This research has made use of NASA's Astrophysics Data System
Bibliographic Services and the NASA/IPAC Extragalactic Database
(NED), which is operated by the Jet Propulsion Laboratory, California
Institute of Technology, under contract with the National Aeronautics
and Space Administration.



\begin{deluxetable}{lcccccccccc}
\tablecaption{ Emission-Line Intensities and Equivalent Widths for Leo~P}
\tablewidth{0pt}
\tablehead{
\CH{} & \multicolumn{2}{c}{$I(\lambda)/I(\mbox{H}\beta)$} }
\startdata
{Ion} 		        	& {KPNO 4-m}   			& {LBT/MODS}   	\\	
\hline																																													
{[O II] $\lambda$3727}		& 0.467 $\pm$ 0.038             & 0.465 $\pm$ 0.017   \\ 
{H12 $\lambda$3750}             & \nodata                       & 0.038 $\pm$ 0.007  \\
{H11 $\lambda$3771}             & 0.040 $\pm$ 0.010             & 0.043 $\pm$ 0.007  \\
{H10 $\lambda$3798}             & 0.045 $\pm$ 0.010             & 0.055 $\pm$ 0.008  \\
{He I $\lambda$3820}		& \nodata        		& 0.010 $\pm$ 0.007  \\
{H9 $\lambda$3835}		& 0.062 $\pm$ 0.009             & 0.072 $\pm$ 0.007   \\	
{[Ne III] $\lambda$3868}	& 0.118 $\pm$ 0.011             & 0.106 $\pm$ 0.008   \\ 
{He I+H8 $\lambda$3889}	        & 0.195 $\pm$ 0.015             & 0.197 $\pm$ 0.009 \\ 
{[Ne III]+H7 $\lambda$3968}	& 0.181 $\pm$ 0.014             & 0.186 $\pm$ 0.009   \\ 
{He I $\lambda$4026}		& \nodata         		& 0.011 $\pm$ 0.007   \\ 
{H$\delta$ $\lambda$4101}	& 0.253 $\pm$ 0.015             & 0.268 $\pm$ 0.009   \\ 
{H$\gamma$ $\lambda$4340}	& 0.473 $\pm$ 0.020             & 0.452 $\pm$ 0.011   \\ 
{[O III] $\lambda$4363}		& 0.035 $\pm$ 0.007             & 0.038 $\pm$ 0.005   \\ 
{He I $\lambda$4471}		& 0.034 $\pm$ 0.005      	& 0.034 $\pm$ 0.006   \\ 
{H$\beta$ $\lambda$4861}	& 1.000 $\pm$ 0.022             & 1.000 $\pm$ 0.020   \\ 	
{He I $\lambda$4921}		& \nodata         		& 0.009 $\pm$ 0.006   \\ 
{[O III] $\lambda$4959}		& 0.468 $\pm$ 0.010             & 0.482 $\pm$ 0.010   \\  
{[O III] $\lambda$5007}		& 1.399 $\pm$ 0.031             & 1.453 $\pm$ 0.030   \\ 
{He I $\lambda$5015}		& \nodata         	        & 0.025 $\pm$ 0.005   \\ 
{He I $\lambda$5876}		& 0.094 $\pm$ 0.006             & 0.098 $\pm$ 0.004   \\ 
{[S III] $\lambda$6312}		& \nodata      	                & 0.007 $\pm$ 0.003   \\	
{[N II] $\lambda$6548}		& 0.008 $\pm$ 0.004		& 0.009 $\pm$ 0.003    \\ 
{H$\alpha$ $\lambda$6563}	& 2.699 $\pm$ 0.197             & 2.750 $\pm$ 0.095   \\ 
{[N II] $\lambda$6584}		& 0.021 $\pm$ 0.005             & 0.025 $\pm$ 0.003    \\ 
{He I $\lambda$6678}		& 0.026 $\pm$ 0.006             & 0.028 $\pm$ 0.003    \\ 
{[S II] $\lambda$6717}		& 0.047 $\pm$ 0.007             & 0.036 $\pm$ 0.003    \\ 
{[S II] $\lambda$6731}		& 0.027 $\pm$ 0.006             & 0.027 $\pm$ 0.003    \\ 
{He I $\lambda$7065}		& \nodata                       & 0.023 $\pm$ 0.002   \\ 
{[Ar III] $\lambda$7136}	& \nodata                       & 0.026 $\pm$ 0.002   \\	
{He I $\lambda$7281}		& \nodata        		& 0.007 $\pm$ 0.002   \\
{[O II] $\lambda$7320}		& \nodata                       & 0.006 $\pm$ 0.002   \\ 
{[O II] $\lambda$7330}		& \nodata                       & 0.007 $\pm$ 0.002   \\ 
{P13 $\lambda$8665}		& \nodata                       & 0.007 $\pm$ 0.002   \\	
{P12 $\lambda$8750}		& \nodata                       & 0.013 $\pm$ 0.002   \\	
{P11 $\lambda$8863}		& \nodata                       & 0.011 $\pm$ 0.002   \\	
{P10 $\lambda$9015}		& \nodata                       & 0.015 $\pm$ 0.003   \\	
{[S III] $\lambda$9069} 	& \nodata                       & 0.043 $\pm$ 0.004   \\	
{P9 $\lambda$9229}		& \nodata                       & 0.020 $\pm$ 0.003   \\	
{[S III] $\lambda$9532} 	& \nodata                       & 0.102 $\pm$ 0.007   \\	
{P8 $\lambda$9546}		& \nodata                       & 0.023 $\pm$ 0.003   \\	
\hline																																													
{C(H$\beta$)}			& 0.00$\pm$0.10  	        & 0.09$\pm$0.04        \\
{EW(H$\beta$(ABS))} (\AA )	& 2.0$\pm$2.0                   & 1.0$\pm$2.0          \\ 
{$F$(H$\beta$)}			& 46$\pm$0.7     	        & 34$\pm$0.7         \\ 
{EW(H$\beta$)} (\AA )		& 143			        & 183                  \\ 
{EW(H$\alpha$)}  (\AA )         & 927  			        & 1222                  \\ 
\hline \\
\enddata
\tablecomments{Emission line fluxes (measured by directly integrating under the line 
profile and only using deblended Gaussian profile fits and multiple component 
fits when necessary) are relative to H$\beta$ = 1.00 and are corrected for reddening. 
The H$\beta$ flux is given for reference, with units of $10^{-16}$ erg s$^{-1}$ cm$^{-2}$. 
Note that uncertainties listed in this table reflect the statistical uncertainties in the 
flux through the slit only, and do not account for slit losses.} 
\label{tbl1}
\end{deluxetable}


\tablenum{2}
\begin{deluxetable}{lrcccc}
\tablewidth{0pt}
\footnotesize
\tablecaption{Ionic and Total Abundances for Leo~P}
\tablehead{
\colhead{Species}      & & \colhead{KPNO 4-m}              & \colhead{LBT/MODS}              & \colhead{Adopted}}
\startdata
\\
T(O$^{++}$)            & & 17150 $^{+2040}_{-1390}$        & 17350 $^{+1390}_{-1060}$        &  \nodata  \\
T(O$^{+}$) (inferred)  & & 14460 $\pm$ 1440                & 14530 $\pm$ 1030                &  \nodata  \\
T(S$^{++}$) (inferred) & & 15930 $\pm$ 1190                & 16010 $\pm$ 1140                &  \nodata  \\
n$_e$(S$^{+}$)         & &  0 $^{+235}_{-0}$ cm$^{-3}$     & 60 $^{+200}_{-60}$ cm$^{-3}$    &  \nodata  \\
n$_e$(O$^{+}$)         & & \nodata                         & 45 $^{+66}_{-45}$ cm$^{-3}$     &  \nodata  \\
\hline
(O$^+$/H)              & & 0.42 $\pm$ 0.13 $\times10^{-5}$ & 0.41 $\pm$ 0.09 $\times10^{-5}$ & \nodata  \\
(O$^{++}$/H)           & & 1.08 $\pm$ 0.22 $\times10^{-5}$ & 1.09 $\pm$ 0.16 $\times10^{-5}$ & \nodata   \\
(O/H)                  & & 1.49 $\pm$ 0.26 $\times10^{-5}$ & 1.50 $\pm$ 0.18 $\times10^{-5}$ & 1.46 $\pm$ 0.14 $\times10^{-5}$ \\
log(O/H)  + 12         & & 7.17 $\pm$ 0.07                 & 7.17 $\pm$ 0.05                 & 7.17 $\pm$ 0.04 \\
\\
(N$^+$/O$^+$)          & & 4.0  $\pm$ 0.9  $\times10^{-2}$ & 4.7  $\pm$ 0.6  $\times10^{-2}$ & 4.4 $\pm$ 0.4 $\times10^{-2}$ \\
log(N/O)               & & $-$1.40  $\pm$ 0.09             & $-$1.33 $\pm$ 0.05              & $-$1.36 $\pm$ 0.04 \\
(N$^+$/H)              & & 1.7  $\pm$ 1.1  $\times10^{-7}$ & 2.0  $\pm$ 0.7  $\times10^{-7}$ & \nodata                         \\
                       & ICF & 3.58 $\pm$ 0.19             & 3.67 $\pm$ 0.13                 & \nodata                         \\
(N/H)                  & & 5.9  $\pm$ 1.7  $\times10^{-7}$ & 7.1  $\pm$ 1.2  $\times10^{-7}$ & 6.7  $\pm$ 1.0  $\times10^{-7}$ \\
log(N/H)  + 12         & & 5.77 $\pm$ 0.11                 & 5.85 $\pm$ 0.07                 & 5.82 $\pm$ 0.06 \\
\\
(Ne$^{++}$/O$^{++}$)   & & 1.92 $\pm$ 0.25 $\times10^{-1}$ & 1.66 $\pm$ 0.14 $\times10^{-1}$ & 1.72 $\pm$ 0.12 $\times10^{-1}$ \\
log(Ne/O)              & & $-$0.72  $\pm$ 0.05             & $-$0.78 $\pm$ 0.04              & $-$0.76 $\pm$ 0.03 \\
(Ne$^{++}$/H)          & & 2.07 $\pm$ 0.56 $\times10^{-6}$ & 1.80 $\pm$ 0.35 $\times10^{-6}$ & \nodata                         \\
                       & ICF & 1.39 $\pm$ 0.19             & 1.37 $\pm$ 0.13                 & \nodata                         \\
(Ne/H)                 & & 2.87 $\pm$ 0.62 $\times10^{-6}$ & 2.48 $\pm$ 0.37 $\times10^{-6}$ & 2.58 $\pm$ 0.32 $\times10^{-6}$ \\
log(Ne/H) + 12         & & 6.46 $\pm$ 0.08                 & 6.39 $\pm$ 0.06                 & 6.42 $\pm$ 0.05 \\
\\
(S$^+$/H)              & & 0.75 $\pm$ 0.25 $\times10^{-7}$ & 0.64 $\pm$ 0.12 $\times10^{-7}$ & \nodata                         \\
(S$^{++}$/H)           & &  \nodata                        & 2.84 $\pm$ 0.33 $\times10^{-7}$ & \nodata                         \\
                       & ICF &  \nodata                    & 1.39 $\pm$ 0.12                 & \nodata                         \\
(S/H)                  & &  \nodata                        & 4.84 $\pm$ 0.55 $\times10^{-7}$ & 4.84 $\pm$ 0.55 $\times10^{-7}$ \\
log(S/H)  + 12         & &   \nodata                       & 5.68 $\pm$ 0.47                 & 5.68 $\pm$ 0.47 \\
log(S/O)               & &  \nodata                        & $-$1.49 $\pm$ 0.07              & $-$1.49 $\pm$ 0.07 \\
\\
(Ar$^{++}$/H)          & & \nodata                         & 9.1 $\pm$ 1.3 $\times10^{-8}$   & \nodata                         \\
                       & ICF & \nodata                     & 1.65 $\pm$ 0.18                 & \nodata                         \\
(Ar/H)                 & & \nodata                         & 1.50 $\pm$ 0.27 $\times10^{-7}$ & 1.50 $\pm$ 0.27 $\times10^{-7}$ \\
log(Ar/H) + 12         & & \nodata                         & 5.18 $\pm$ 0.07                 & 5.18 $\pm$ 0.07 \\
log(Ar/O)              & &  \nodata                        & $-$2.00 $\pm$ 0.09              & $-$2.00 $\pm$ 0.09 \\
\enddata
\label{tbl2}
\end{deluxetable}


\tablenum{3}
\begin{deluxetable}{lcccc}
\tablecaption{Inputs and $\chi ^2$s for MCMC Analysis of Helium Abundance}
\tablewidth{0pt}
\tablehead{
\colhead{Emission Line}        & \colhead{Flux}         & \colhead{EW}   & \colhead{$\chi ^2$} & \colhead{$+$ $\lambda$5015 $\chi ^2$} }
\startdata

{H$\delta$ $\lambda$4101}       & 0.246 $\pm$ 0.006     & 26.5  $\pm$ 2.7   &   0.252   & 0.310 \\    
{H$\gamma$ $\lambda$4340}       & 0.431 $\pm$ 0.009     & 54.6  $\pm$ 5.5   &   2.751   & 2.686 \\   
{H$\beta$ $\lambda$4861}        & 1.000 $\pm$ 0.020     & 183.2 $\pm$ 18.3  &  \nodata  & \nodata  \\
{H$\alpha$ $\lambda$6563}	& 2.949 $\pm$ 0.059     & 1222  $\pm$ 122   &   0.191   & 0.177 \\ 
{He I+H8 $\lambda$3889}	        & 0.175 $\pm$ 0.007     & 16.3  $\pm$ 1.6   &   0.019   & 0.009 \\ 
{He I $\lambda$4026}		& 0.011 $\pm$ 0.007    	& 1.0   $\pm$ 0.1   &   0.004   & 0.029   \\ 
{He I $\lambda$4471}		& 0.033 $\pm$ 0.006    	& 4.5   $\pm$ 0.5   &   0.015   & 0.034 \\ 
{He I $\lambda$5015}		& 0.026 $\pm$ 0.006     & 5.0   $\pm$ 0.5   &  \nodata  & 0.211 \\ 
{He I $\lambda$5876}		& 0.103 $\pm$ 0.004     & 30.2  $\pm$ 3.0   &   0.001   & 0.002  \\ 
{He I $\lambda$6678}		& 0.030 $\pm$ 0.003     & 14.3  $\pm$ 1.4   &   0.024   & 0.022  \\ 
{He I $\lambda$7065}		& 0.025 $\pm$ 0.003     & 12.8  $\pm$ 1.3   &   0.028   & 0.044  \\ 
\hline
Total $\chi$$^2$                &  \nodata              &   \nodata         &  3.29     & 3.50     \\
\enddata
\tablecomments{Emission line fluxes are relative to H$\beta$ = 1.00 and are {\it not} corrected for reddening
(as reddening is derived in the MCMC analysis).}
\label{tbl3}
\end{deluxetable}


\tablenum{4}
\begin{deluxetable}{lccc}
\tablecaption{MCMC Analysis of Helium Abundances and Physical Conditions for Leo~P}
\tablewidth{0pt}
\tablehead{
\colhead{Parameter}   & \colhead{Standard}            & \colhead{$+$ $\lambda$5015} }
\startdata
Emission lines        &  9                            & 10                             \\
Free Parameters       &  8                            &  8                             \\
(He$^+$/H$^+$)        &  0.0837$^{+0.0084}_{-0.0062}$ & 0.0837$^{+0.0084}_{-0.0054}$   \\
T$_e$ (10$^4$K)       &  1.71$^{+0.19}_{-0.28}$       &  1.72$^{+0.18}_{-0.27}$        \\
n$_e$ (cm$^{-3}$)     &  1$^{+206}_{-1}$              &  1$^{+201}_{-1}$               \\
C(H$\beta$)           &  0.10$^{+0.03}_{-0.07}$       &  0.10$^{+0.03}_{-0.07}$        \\
a$_H$ (EW \AA)        &  0.94$^{+1.44}_{-0.94}$       &  1.02$^{+1.43}_{-1.02}$        \\
a$_{He}$ (EW \AA)     &  0.50$^{+0.42}_{-0.42}$       &  0.45$^{+0.39}_{-0.45}$        \\
$\tau$$_{He}$         &  0.00$^{+0.66}_{-0.00}$       &  0.00$^{+0.65}_{-0.00}$        \\
n$_{HI}$ (10$^{-4}$ cm$^{-3}$)  & 0$^{+156}_{-0}$     & 0$^{+152}_{-0}$                \\
$\chi$$^2$            &  3.3                          & 3.5                            \\
$\chi$$^2$/d.o.f.     &  3.3                          & 1.75                           \\
Y                     & 0.2509$^{+0.0184}_{-0.0142}$  & 0.2509$^{+0.0184}_{-0.0123}$   \\
\enddata
\label{tbl4}
\end{deluxetable}


\end{document}